# Atomic-Scale Visualization of Chiral Charge Density Wave States and Their Reversible Transition

Xuan Song[1#], Liwei Liu[1#*], Yaoyao Chen[1], Han Yang[1], Zeping Huang[1], Baofei Hou[1], Yanhui Hou[1], Xu Han[1], Huixia Yang[1], Quanzhen Zhang[1], Teng Zhang[1], Jiadong Zhou[1], Yuan Huang[1], Yu Zhang[1], Hong-Jun Gao[2] and Yeliang Wang[1*]

[1]School of Integrated Circuits and Electronics, MIIT Key Laboratory for Low-Dimensional Quantum Structure and Devices, Beijing Institute of Technology, Beijing 100081, China.
[2]Institute of Physics, Chinese Academy of Sciences, Beijing 100190, China.

**Chirality is essential for various amazing phenomena in life and matter. However, chirality and its switching in electronic superlattices, such as charge density wave (CDW) arrays, remain elusive. In this study, we characterize the chirality transition with atom-resolution imaging in a single-layer $NbSe_2$ CDW pattern by technique of scanning tunneling microscopy. The atomic lattice of the CDW array is found continuous and intact although its chirality is switched. Several intermediate states are tracked by time-resolved imaging, revealing the fast and dynamic chirality transition. Importantly, the switching is reversibly realized with an external electric-field. Our findings unveil the delicate transition process of chiral CDW array in a 2D crystal down to the atomic scale and may be applicable for future nanoscale devices.**

Chirality refers to the character of an object or state that cannot be superimposed on its mirror image[1,2], which can be used to tune various intriguing properties in condensed matter, such as p-wave superconductors[3,4], solitons[5,6], and Luttinger liquids[7]. However, chirality and its switching in CDW systems, as amazing electronic superlattices with potential applications in ultrafast switches and memories, have rarely been reported. So far, only under specific conditions, such as those in doped TaSe$_2$[8] or TaS$_2$[9,10], laser-pulse-excited TaS$_2$[11], and voltage-pulse treated 2H-TaSe$_2$[12], the coexistence of chiral CDW domains have been observed. However, it is still a big challenge to characterize the chiral domains and the dynamic transition process at the atomic level, as well as realize reversible chirality switch in CDW systems.

In this study, we used the molecular beam epitaxy (MBE) method to grow high-quality single-layer (SL) 1T-NbSe$_2$ islands on bilayer graphene (BLG)[13]. We demonstrate that the SL 1T-NbSe$_2$, a classic 2D CDW crystal, has intrinsic coexisting of chiral domains, which provides a versatile platform to study the nature of chiral CDWs and the switching process. Notably, using scanning tunneling microscopy (STM), we achieve the atomic-resolution imaging of the chiral CDW patterns, and reveal that the atomic lattice is continuous before and after the chirality switch. Furthermore, chirality switch of CDW domains, is reversibly manipulated by the external electric field. Our findings provide novel perspectives for chiral CDW manipulation under the atomic scale.

**Atomic-scale characterization of chiral CDW patterns of SL 1T-NbSe$_2$**

Figure 1 shows the chiral CDW patterns in an SL 1T-NbSe$_2$. The atomic model of SL 1T-NbSe$_2$ and its star of David (SOD) unit is displayed in Fig. 1a. In the 1T phase, each Nb atom is sandwiched by six Se atoms in an octahedral configuration, forming a Se-Nb-Se structure[14]. Each SOD cluster contains thirteen Nb atoms in the middle plane, six Se atoms in the top plane, and the other six in the bottom plane[15]. In the STM topography (Fig. 1b), the top-layer Se atoms contribute more and is usually the only visible layer. Each SOD is imaged as a triangle, due to partial Se atoms in the top layer having more contribution to the apparent contrast[13].

The magnified STM images in Fig. 1c reveal the atom-resolved lattice of the top Se layer in the SOD array. Note that although a single triangular-shaped SOD looks similar over the whole island, there are two chiral domains with different SOD arrangements, as labeled by the green and blue triangles. Compared to the Se atomic close-packed direction (indicated by the black vertical dashed-line), the directions of the SOD triangles in the two domains (indicated by the blue and green arrows) are rotated by around ± 14°, respectively.

The rotation of 14° can be either counterclockwise (CCW, L chirality), or clockwise (CW, R chirality) and should be energetically degenerate[11]. As shown by the model in Fig. 1d, the basis vectors of the chiral CDW domains are not equivalent: $b_{L1}$ and $b_{L2}$ in the green rhombus for L domain, and $b_{R1}$ and $b_{R2}$ for the blue rhombus for the R domain, respectively. Accordingly, the chirality of CDW domains can be defined as "left-handed" (L) and "right-handed" (R), respectively.

Here, two approaches were used to identify the chiralities of the CDW patterns. One is by the ±14° rotation between the close-packed direction of the CDW superlattice and that of the top Se atomic lattice (see the green, blue and black arrows in Fig. 1c). The other approach is by the relative arrangement of the three-neighboring SODs clusters. As shown by the blue or green windwheels in Fig. 1c, the different interlocking manners of the triangular-shaped SOD clusters can be used to distinguish their chiralities from each other, even if the STM images do not have the atomic resolution. Such chirality can be intuitively understood because the SOD superlattice has a ($\sqrt{13} \times \sqrt{13}$)R13.9° relationship with respect to the top Se atomic lattice[16].

**Visualization of atomic lattice before and after chiral CDW chirality transition**

Next, we concentrate on the chirality change, and we find that the atomic lattice of the CDW pattern is still continuous and intact after the chirality change. Figure 2a-c shows the movement of a domain boundary (DB) in the CDW patterns. The DB moves rightward from the yellow dashed-line (Fig. 2a) to the red one (Fig. 2b), corresponding to the expansion of the L domain and shrinking of the R domain. It further transforms into one domain with a single L chirality, as shown in Fig. 2c. To better illustrate this chiral CDW transition process at the atomic scale, we superpose the models of both the SOD arrays and atomic lattice onto the STM images, as shown in Fig. 2d-f. The green and blue balls represent the top-layer Se atoms within the SOD clusters with left and right chiralities, respectively, while the gray balls represent the Se atoms out of the SOD clusters. The STM images of Fig. 2a-c without color-coding are also shown in Supplementary Fig. S1 for comparison.

Importantly, two features remain unchanged after the DB movement of the chiral CDW patterns, as exemplified in Fig. 2a-f: (1) The well-ordered hexagonal array of balls which describes the positions of the top Se lattice, is continuous not only at each single chiral CDW domain but also across the DB. Namely, the top-layer Se atoms keep the same periodicity with the continuous arrangement. (2) The orientations of all triangular SOD at the two chiral CDW domains keep the same. These two features together suggest that the whole Se-Nb-Se structure is preserved over the island, with the continuous 1T structure in the two chiral CDW domains. These features are also illustrated in the side view in Supplementary Fig. S1.

To the best of our knowledge, this is the first *in-situ* atomic-resolution analysis of the DB movement in 2D chiral CDW materials. The intact and robust atomic structure is an important virtue for future chiral-CDW based devices, as it avoids the atomic defect at the DB and will be favorable for the reversible switch application. More examples of the DB movement of the chiral CDW patterns are shown in Supplementary Fig. S2 and Supplementary Movie. Notably, although there are some "imperfect SOD" at the DB (labeled by both green and blue balls), the intact and robust atomic structure is still preserved.

These geometric features obtained by the STM images are also reflected in the FFT images (Fig. 2g-i). All FFT images reveal a single set of reciprocal spots (as marked by the outer brown circles in Fig. 2g-i, labeled as $a_1^*$ and $a_2^*$), which originated from the well-ordered Se lattice. In contrast, another two sets of spots (marked by green and blue circles, respectively) with mirror symmetry are shown in Fig. 2g,h, which can be ascribed to the chiral CDW patterns. The intensities of the two sets of spots show a change following the shift of chiral domain boundary. The spots marked by green circles have larger intensity, corresponding to expanding of the L domain in the real space (Fig. 2a,b). The spots in Fig. 2i are reduced to one set, corresponding to the single chiral domain (Fig. 2f). This kind of intensity change is also illustrated clearly by the magnified FFT image (Supplementary Fig. S3).

**Tracking the dynamic process in chiral CDW patterns**

After resolving the atomic arrangement with the ultrahigh spatial resolution, we further investigated the dynamic process of chiral change in CDW patterns with high time resolution. Two metastable statuses at the chiral DB are tracked by STM observations, as shown by status I and II in Fig. 3a,b, respectively. The fuzzy SOD sites (denoted by dashed triangles) at the DB belong to either (or both) L or R domain. The transition rate is so high that instant SODs are different to capture by normal STM scanning (500 ms/scan line) and thus fuzzy features are present. These fuzzy SOD sites look like touching or overlapping, different from the ideal SOD array with hexagonal lattice. Taking the region around one fuzzy SOD cluster in the 3D STM images (Fig. 3d,e) as an example, the SOD cluster has a change of the residence time at the L and R domains, respectively.

To track the dynamic process of the chiral domain transition, we performed current–time ($I$–$t$) spectroscopy (0.05 ms/point) by holding the STM tip over a fuzzy region (marked by the crosses in Fig. 3a,b and Fig. 3d,e) with a constant height mode. Such $I$–$t$ measurement is a reliable STM method described in the literature for studying the dynamic process of on-surface adsorbates, such as tautomerization[17,18], hopping or rotation[19-22], bond breaking[23], and even orbital dynamics[24]. In our cases, when the fluctuating SOD site is just beneath the tip (e.g., at the L domain in Fig. 3a,d), the tip-sample distance is smaller and the local density of states is enhanced, thus the tunneling current is higher. When the fluctuating SOD is away from the tip (e.g., in the R domain in Fig. 3b,e), the tip-sample distance enlarges and the local density of states reduces; thus, the current will be lower.

In this way, the low and high current values are obtained following the SOD occupation change in two chiral domains, which provide direct evidence of the fast dynamic process of the chiral change. The current-time (*I–t*) curve in Fig. 3f shows the evolution of the current in the duration of 40 ms. The low state is at 100 pA, and it changes to a high state at 240 pA for approximately 2.5 ms, and then changes back.

By increasing the tunneling current, the transition events become more frequent (Fig. 3g for the *I-t* curve with a setpoint of 400 pA). The statistical analysis of the residence time distribution (displayed in Supplementary Fig. S4) reveals that it can be fitted by an exponential curve ($y = y_0 + Ae^{Rx}$, where the transition rate $R$ is determined by the fitting, $y_0$ and $A$ are constants). By plotting the transition rate against the tunneling current with a *log* scale, a linear relationship with a slope of 1.042 ± 0.055 is obtained (Fig. 3h), indicating a one-electron process[17,18,25] during the chiral CDW transition. Thus, the transition rate can be tuned by the tunneling current.

Based on these experimental findings, the mechanism of the chiral CDW transition process is illustrated by the schematic in Fig. 3i. Regarding the intermediate CDW statues I and II, they jump fast over the low energy barrier, back and forth. When the intermediate state obtains sufficient energy to jump over the higher barrier, it falls into more stable status without fuzzy features (status III). In short, using the current-time curve to monitor the dynamic transition process of chiral CDW patterns, we observed several intermediate states at the chiral DB. As far as we know, this is the first dynamic study together with the atomic characterization on a chiral 2D CDW system.

**Reversible switch of chiral CDW patterns induced by an external electric field**

After visualizing the switching process of the chirality at the atomic scale, we further managed to switch the chiral CDW islands as a whole reversibly, as shown in Fig. 4a-c. A schematic of the experimental setup is shown in Fig. 4g. The switching process from L to R chirality was realized by applying voltage pulses of ~ -5.5 V at a lateral distance of approximately 5 nm from the STM tip to the edge of the 1T-NbSe$_2$ island. The reversible switching process from R to L chirality was realized by applying voltage pulses of ~ -6.6 V. Taking advantage of our simultaneous imaging of chiral CDW patterns with atomic resolution (Supplementary Fig. S5), we demonstrate that the atomic lattice is still continuous and intact although the chirality of CDW is switched.

Note that even if the SOD protrusions are not atomically resolved in the large-scale STM images, the chiralities and chiral change of the CDW domains can still be identified in another easy way. As shown in Fig. 4a-c, the SOD close-packed directions have rotation angles of ±14º compared to the direction of the island edge, with +14º corresponding to L and -14º corresponding to R, similar to those defined in Fig. 1c-d. The magnified images of the triangular SOD arrays of the R and L domains are shown in Fig. 4d-f, also demonstrating the chirality change well.

As far as we tested, at a lateral distance of ~20 nm, the chirality can still be switched with a voltage-pulse of 10 V. In addition, the chirality of the island of 50 nm can be switched as a whole, suggesting a domino-like effect in changing the chirality. This effect should be associated with the collective electronic nature of the CDW. Notably, the CDW chirality of the islands can be stable even above room temperature (RT) with small islands size of 15 nm. This advantage further facilitates the application of chiral-CDW based nanodevices above RT.

In conclusion, with high-spatial-resolution imaging, fast-tracking $I$–$t$ spectroscopy, and voltage-pulse manipulation, we have comprehensively investigated the atomic structure, dynamic process, and reversible manipulation of the chiral 2D CDW patterns for the first time, in a 2D crystal of SL 1T-NbSe$_2$. In our study, three important features are identified: (1) The top-layer Se atom lattice of NbSe$_2$ is unchanged although the CDW chirality changed. (2) Chirality transition of CDW patterns is fast and intermediate states exist in the dynamic transition process. (3) The chiral switching can be reversibly manipulated in the CDW domains, induced by an external electric-field. Our findings provide in-depth understandings of the structure and manipulation of chiral CDW systems down to the atomic level. The 2D chiral CDW systems may lay a versatile stage for both fundamental understanding and practical application in future nanoelectronics and nanotechnology.

**Supplementary Information** is available in the online version of the paper.

**Acknowledgments**

This work was supported by the National Key Research and Development Program of China (2020YFA0308800, 2019YFA0308000), National Natural Science Foundation of China (Nos. 61971035, 61725107, 61901038), Beijing Natural Science Foundation (Nos. Z190006, 4192054).

**Author Contributions**

L.W.L., H.J.G. and Y.L.W. coordinated the research project. X.S., L.W.L., Y.Y.C., H.Y., Z.P.H., B.F.H., Y.H.H, H.X.Y, J.D.Z, T.Z., Y.Z. and Y.L.W. performed the experimental fabrication and measurements. All authors analyzed the data and discussed the manuscript. X.S. and L.W.L. made equal contributions to the work.
**Competing interests** The authors declare no competing financial interests.

**Additional Information**
**Reprints and permissions information** is available at www.nature.com/reprints. Readers are welcome to comment on the online version of the paper.
**Correspondence and requests for materials** should be addressed to L.W.L. (liwei.liu@bit.edu.cn) and Y.L.W. (yeliang.wang@bit.edu.cn).

**Methods**

Sample preparation and STM measurements. MBE and STM experiments were performed by a Unisoku ultrahigh-vacuum system (base pressure ~1 × $10^{-10}$ mbar) comprising an MBE chamber and LT-STM chamber. Epitaxial graphene was grown by thermal decomposition of 4H-SiC(0001) at 1200 °C for 40 min. NbSe$_2$ islands were prepared using the MBE method at a deposition rate of ~0.002 ML/min. An elemental Nb rod was used as the metal source in the e-beam evaporators (Focus Ltd.), whereas an elemental Se source in a Knudsen cell was heated to 120 °C. The deposition ratio of Se to Nb was larger than 20:1. During the film deposition, the background pressure in the chamber was ~3.0 × 10$^{-9}$ mbar. The BLG/SiC substrate was stored at ~550 °C during the deposition and post-annealing processes to improve the diffusion and desorb excess Se. STM imaging was performed using mechanically cut Pt-Ir tips at the liquid helium temperature (4.2 K). The d$I$/d$V$ tunneling spectra were acquired using lock-in detection by applying an alternating-current modulation of 20 mV (root mean square) at 973 Hz to the bias voltage.

**Data availability**. The main data supporting the findings of this study are available within the article, and Supplementary Information. Extra data are available from the corresponding author upon reasonable request.

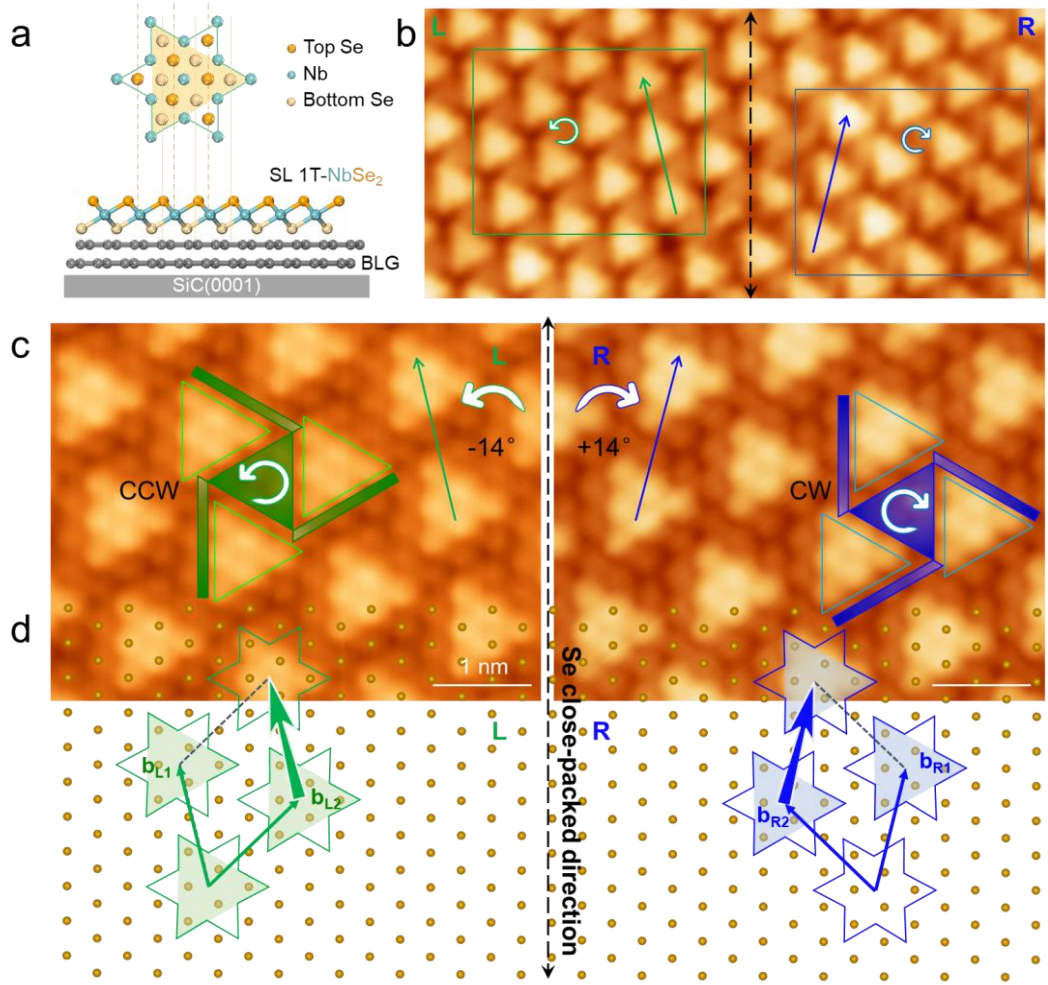

**Figure 1 | STM topography and schematic model of chiral CDW patterns. a**, (upper panel) top view of SOD model showing thirteen Nb atoms, six top Se atoms, and six bottom Se atoms, with three-fold symmetry; (lower panel) side view of SL 1T-NbSe$_2$ on a BLG/SiC(0001) substrate. The top- and bottom-layer Se atoms are projected from the upper to the lower panel to show the relative positions. **b**, CDW-resolved STM topography around a domain boundary of SL 1T-NbSe$_2$. The top-Se close-packed direction is marked by the black dashed arrow. **c**, Left (right): Zoomed-in atomic-resolved image in the L (R) domain marked by the green (blue) frame in **b**. The SOD arrays in these two domains are labeled by green and blue triangles, respectively. The SODs arrangements are represented by the clockwise (CW) and counterclockwise (CCW) arrows. **d**, Atomic model of SL 1T-NbSe$_2$ with chiral CDW domains. The atomic lattice of the top layer Se is indicated by the array of orange balls, and the SOD clusters in these two domains are labeled by green and blue triangles encircled by

hexagram stars, respectively. Green and blue arrows crossing the adjacent SOD triangles represent the chiral CDW close-packed directions, which are rotated by ± 14° with respect to the Se atom close-packed direction. Scanning parameters: **b**, bias voltage $V_B = -1.5$ V, tunneling current $I_t = 1$ nA; **c**, $V_B = -1.5$ V, $I_t = 2$ nA.

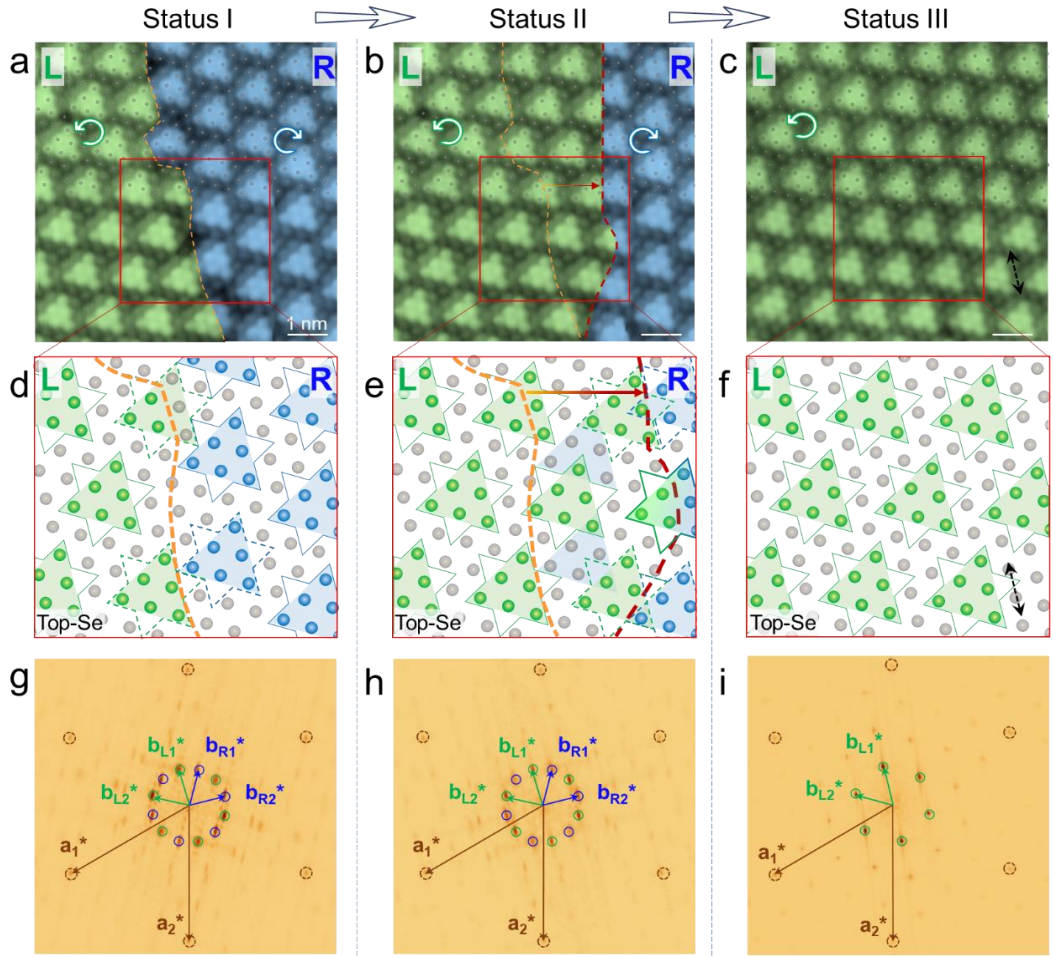

**Figure 2 | Continuous atomic lattice during the chirality transition of the CDW pattern. a-c**, Atomic-resolution images of the same area of the 1T-NbSe$_2$ island before (**a**) and after (**b**) boundary movement of chiral domains, demonstrating a chiral change from a coexistence of the L-R domain to a pure L domain (**c**). The upper part of each image is overlaid by a well-ordered lattice of gray balls, indicating that the top-layer Se lattice is continuous over the whole area from L to R domains. **d-f**, Atomic model of SL 1T-NbSe$_2$ with the shifted CDW DB, describing the same area marked by the red

frame in **a-c**. The green and blue balls represent Se atoms in the SOD with left and right chiralities, respectively, while the gray balls show the Se atoms surrounding the SOD. All Se atoms (in blue, green, and grey) share the same periodic hexagonal lattice. **g-i**, FFT images of **a-c** showing that both statuses I and II have two chiral sets of spots. In status III, there is a reduction to one set of spots for the CDW pattern. Scanning parameters: **a-c**, −1.5 V, 1 nA.

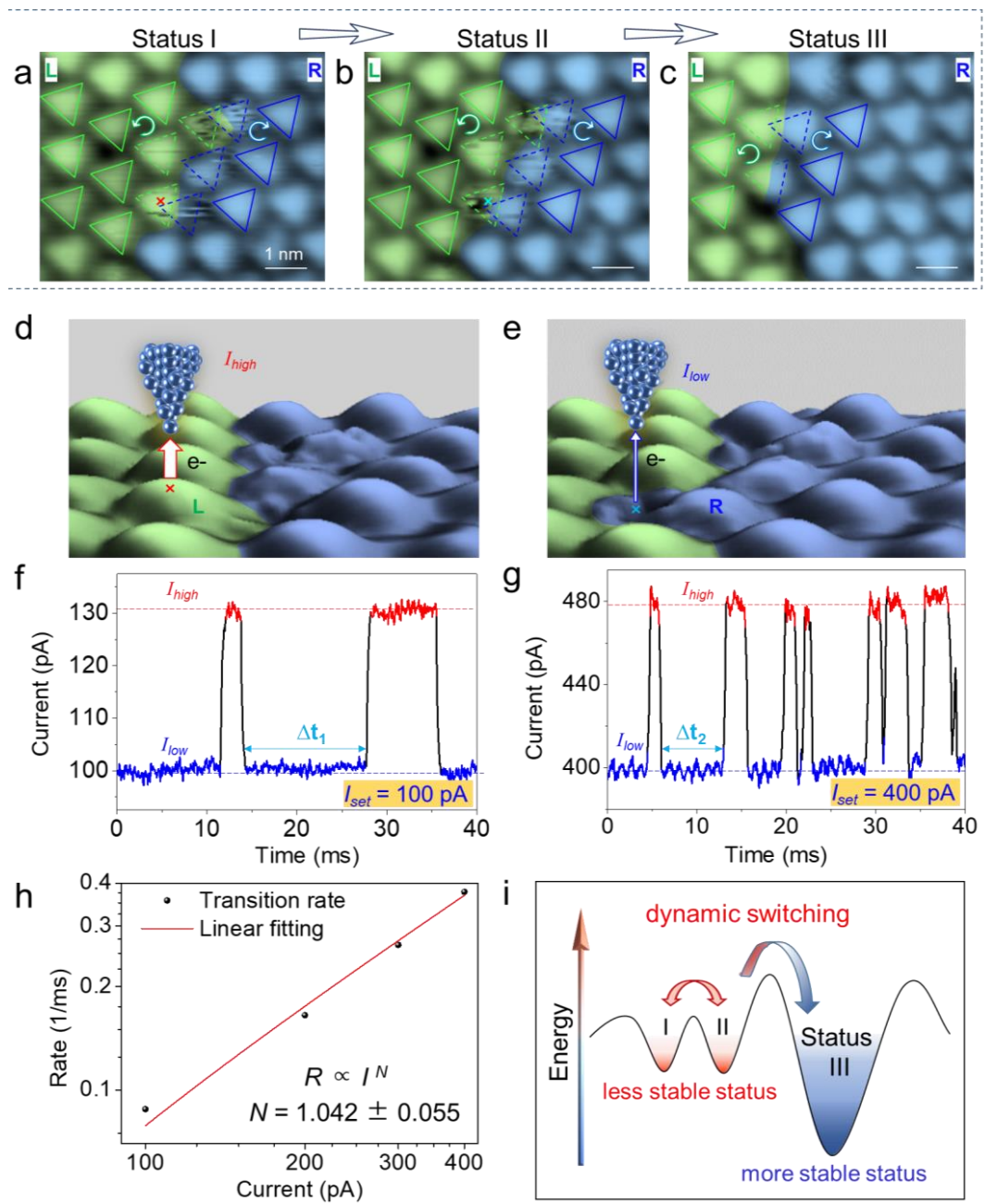

**Figure 3 | Dynamic process of the chirality switching in the CDW superlattice. a** and **b**, STM topography of the same area in a 1T-NbSe$_2$ island, showing the fast switching between transition states I and II around the chiral DB. Green and blue triangular arrays are superimposed on the STM images and indicate the L and R domains, as well as the configuration at the chiral DB. The dashed SODs refer to the CDW sites, which are fuzzy due to the fast switching between two chiral domains. **c**, State III showing the same area as in **a** and **b** with the stabilized chiral DB. **d** and **e**, Schematic of the STM tip above the SOD at the fuzzy chiral DB (3D views of **a** and **b**) to monitor the chiral change, which results in high and low tunneling currents ($I_{high}$ and $I_{low}$), respectively. **f** and **g**, $I-t$ curve recorded with gaps of −1.5 V, 100 pA before turning off the FB loop and −1.5 V, 400 pA, respectively. The transition rate of the fuzzy chiral SOD is tuned by the tunneling current. **h**, Transition rate ($R$) as a function of the tunneling current ($I$) plotted in the *log* scale. The linear line represents the power-law fit ($R \propto I^N$) with $N = 1.042 \pm 0.055$ indicating a one-electron process. **i**, Schematic of chiral CDW changes induced by energy. Scanning parameters: **a**–**c**, −1.5 V, 200 pA.

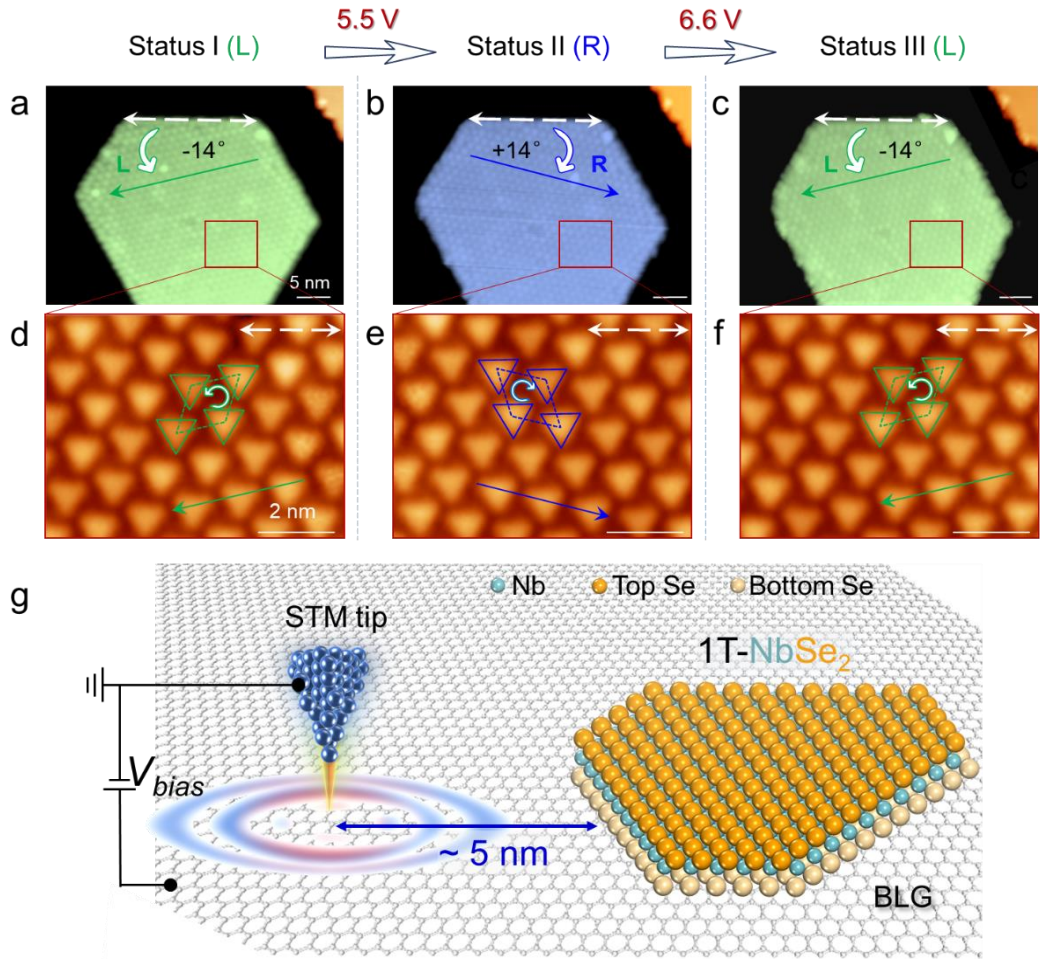

**Figure 4 | Reversible switching of chiral CDW patterns induced by an external electric field. a-c**, Large-scale STM images of the reversible switching of chiral CDW patterns of a 1T-NbSe$_2$ island, realized by applying voltage pulses. Two CDW close-packed directions are represented by green and blue arrows, respectively, which are rotated around ±14° with respect to the island edges (labeled by the white dashed double arrows). **d-f**, Magnified images of the areas in **a-c**, showing the chirality identified by the relative arrangement of neighboring triangular SOD clusters. The top-layer Se close-packed direction is marked by the black dashed-arrow. **g**, Schematic of the experimental setup. The bias voltage is applied between the tip and the substrate to form an external electric-field.